\documentclass[letterpaper,showpacs,preprintnumbers,amsmath,amssymb,nofootinbib,twocolumn,superscriptaddress,notitlepage,prl]{revtex4-1}

\usepackage[usenames,dvipsnames]{color}
\usepackage[colorlinks,citecolor=blue]{hyperref}
\usepackage{amsmath}
\usepackage{bbm}
\usepackage{amsfonts}
\usepackage{amssymb}
\usepackage{latexsym}
\usepackage{graphicx}
\usepackage[english]{babel}
\usepackage{multirow}
\usepackage{float}
\usepackage{url}
\usepackage{slashed}
\usepackage{xcolor} 

\newcommand{\be}{\begin{equation}}
\newcommand{\ee}{\end{equation}}
\newcommand{\ba}{\begin{array}}
\newcommand{\ea}{\end{array}}
\newcommand{\bea}{\begin{eqnarray}}
\newcommand{\eea}{\end{eqnarray}}

\newcommand{\eesasaa}{$e^+ e^- \to \slashed{\gamma}  \slashed{\gamma} \gamma$\,}
\newcommand{\eesesea}{$e^+ e^- \to \slashed{e}^+  \slashed{e}^- \gamma$\,}

\newcommand{\eeslsla}{$e^+ e^- \to \slashed{\ell}^+  \slashed{\ell}^- \gamma$\,}

\begin{document}
\title{Millicharged particles at electron colliders}

\author{Jinhan Liang} 
\affiliation{Department of Physics, Nanjing University, Nanjing 210093, China}

\author{Zuowei Liu} 
\affiliation{Department of Physics, Nanjing University, Nanjing 210093, China} 
\affiliation{Center for High Energy Physics, Peking University, Beijing 100871, China} 
\affiliation{Nanjing Proton Source Research and Design Center, Nanjing 210093, China} 
\affiliation{CAS Center for Excellence in Particle Physics, Beijing 100049, China} 

\author{Yue Ma} 
\affiliation{Kuang Yaming Honors School, Nanjing University, Nanjing 210023, China}

\author{Yu Zhang} 
\affiliation{Institutes of Physical Science and Information Technology, Anhui University, Hefei 230601, China}
\affiliation{School of Physics and Materials Science, Anhui University, Hefei 230601,China}

\begin{abstract}

We propose to search for millicharged particles 
in electron colliders operated with the 
center-of-mass energies at ${\cal O}$(1-10) GeV, 
which include Belle II, BESIII, BaBar, 
and also the proposed experiment STCF. 
We use the monophoton final state 
at electron colliders
to probe the 
parameter space of  
millicharged particles,  
that is spanned by millicharge $\epsilon$ and mass $m$.
We find that electron colliders have sensitivity to the 
previously unexplored parameter space 
for millicharged particles with MeV-GeV mass: 
$\epsilon \lesssim {\cal O}(10^{-1})$ 
for $0.5$ GeV $\lesssim m \lesssim 3.5$ GeV in BaBar, 
$\epsilon \lesssim {\cal O}(10^{-3})$ for  $0.1$ GeV $\lesssim m \lesssim 1.5$ GeV in BESIII,  
$\epsilon \lesssim 10^{-3}-10^{-2}$ for $0.1$ GeV $\lesssim m \lesssim 4$ GeV in Belle II, 
and $\epsilon \lesssim {\cal O}(10^{-4})$ for $1$ MeV $\lesssim m \lesssim 1$ GeV in STCF. 

\end{abstract}

\maketitle

\section{Introduction}

Although anomaly cancellations link the electric charges of  
the standard model (SM) fermions \cite{Geng:1988pr}, 
in principle, there is no such constraint for 
particles beyond the SM (BSM). 
For example, particles with arbitrarily small electric 
charge can naturally arise in models where hidden sectors 
particles interact with the SM particles 
via kinetic mixing 
\cite{Holdom:1985ag, Holdom:1986eq, Foot:1991kb}, 
or via Stueckelberg mixing 
\cite{Kors:2004dx, Cheung:2007ut, Feldman:2007wj}. 
A variety of experiments and theoretical investigations 
have been carried out to search for 
BSM particles with electric charge significantly 
smaller than the electron, which we refer to as millicharged 
particles (MCPs). 
The constraints on MCPs 
come both from terrestrial particle accelerators  
and from astrophysical processes.
Previous particle accelerator constraints on MCPs include 
colliders \cite{Davidson:1991si, Davidson:2000hf, CMS:2012xi} , 
SLAC electron beam dump experiment \cite{Prinz:1998ua}, 
and E613 \cite{Golowich:1986tj, Soper:2014ska}. 
Recently sensitivity of probing MCPs has also been studied  
in various accelerator experiments, including 
BESIII \cite{Liu:2018jdi}, 
LHC \cite{Haas:2014dda}, 
Circular Electron Positron Collider (CEPC) \cite{Liu:2019ogn}, 
NA64 \cite{Gninenko:2018ter, Chu:2018qrm}, 
and Light Dark Matter eXperiment (LDMX) \cite{Berlin:2018bsc}. 
Astrophysical constraints include 
white dwarf \cite{Dobroliubov:1989mr, Davidson:1991si, Davidson:2000hf}, 
supernova \cite{Mohapatra:1990vq, Davidson:2000hf, Chang:2018rso}, 
cosmic microwave background (CMB) \cite{Dubovsky:2003yn, Dolgov:2013una}, 
big bang nucleosynthesis
\cite{Davidson:1991si, Davidson:1993sj, Davidson:2000hf, Vogel:2013raa, Vinyoles:2015khy}, 
red giants 
\cite{Dobroliubov:1989mr, Davidson:1991si, Davidson:1993sj, Vogel:2013raa}, 
and Sun \cite{Vinyoles:2015khy}. 
MCPs can also be searched for in various neutrino experiments 
\cite{Gninenko:2006fi, Magill:2018tbb, Kelly:2018brz, Singh:2018von, Harnik:2019zee}.

Recently, the 21 cm signal  
measured by the Experiment to Detect the Global Epoch of reionization Signature (EDGES)  indicates  
that the universe is colder than expected during the 
cosmic dawn \cite{Bowman:2018yin}. 
Millicharged dark matter (DM) can provide 
cooling to the cosmic hydrogens leading to 
the strong 21 cm absorption signal 
\cite{Munoz:2018pzp, Barkana:2018cct, Berlin:2018sjs, 
Kovetz:2018zan, Boddy:2018wzy, Klop:2018ltd, 
Creque-Sarbinowski:2019mcm, Liu:2019knx}.

In this paper, we study the experimental sensitivity on MCPs  
from electron colliders. 
The constraint on MCPs from the BESIII experiment 
has been recently studied in \cite{Liu:2018jdi}; 
here we extend the analysis to other electron 
colliders operated the GeV scale, 
including Belle II, BaBar, 
and also the proposed experiment, 
the Super Tau Charm Factory (STCF). 
Unlike the DM constraints which assume a sufficient 
amount of millicharged DM in our universe, 
particle colliders can provide robust constraints 
on the MCPs which is independent on its composition 
in the universe. 
At the MeV-GeV scale, the leading constraints 
on MCPs come from 
colliders \cite{Davidson:1991si}, 
SLAC \cite{Prinz:1998ua}, 
and Liquid Scintillator Neutrino Detector (LSND)/MiniBooNE \cite{Magill:2018tbb}. 
We find 
that electron colliders can 
probe the previously unexplored MCP parameter space 
with MeV-GeV mass. 
Our analysis also has a direct impact on 
millicharged DM models that are invoked to 
explain the 21 cm anomaly.

\section{Electron collider signals}

In our analysis, we assume that the MCP is a Dirac fermion 
which is charged under the SM photon via 
the interaction Lagrangian, 
$
\mathcal{L}_{\mathrm{int}}=e \varepsilon A_{\mu} \bar{\chi} \gamma^{\mu} \chi,
$
where $\chi$ is the MCP, $A_\mu$ is the SM photon. 
The analysis presented here can be easily extended to 
MCPs with other spins.

\begin{figure}[htbp]
\begin{center}
\includegraphics[width=0.18\textwidth]{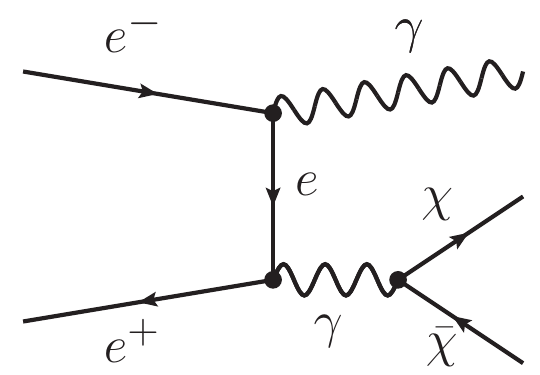}
\hspace{0.2cm}
\includegraphics[width=0.18\textwidth]{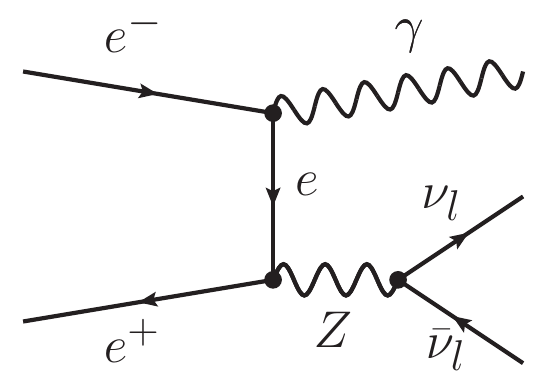}
\caption{Feynman diagrams for the process 
$e^+ e^- \rightarrow \chi \bar{\chi} \gamma$ (left) 
and 
$e^+ e^- \rightarrow \nu \bar \nu \gamma$ (right).}
\label{F-FeynDia-eeXXA}
\end{center}
\end{figure}

Because the ionization signal of the MCP 
is typically undetectable in collider experiments, 
one thus relies on the visible final state particles 
produced in association with MCPs 
for the detection. Thus we use the mono-photon final state 
in electron colliders to search 
for MCPs \cite{Liu:2018jdi}. 
The Feynman diagram 
for the signal process is 
shown in Fig.\ (\ref{F-FeynDia-eeXXA}). 
The maximum photon energy is 
$E^{\rm max}_\gamma = (s-4m_\chi^2)/(2\sqrt{s})$,
which applies to all detector cuts throughout 
this analysis. 
Here $m_\chi$ is the MCP mass 
and $s$ is the square of center-of-mass energy. 

Belle II is operated on SuperKEKB 
which collides 7 GeV electrons with 4 GeV positrons \cite{Kou:2018nap}. 
SuperKEKB has a design luminosity of
$8\times 10^{35}$ cm$^{-2}$ s$^{-1}$ 
and expects to collect 50 $\rm{ab}^{-1}$ integrated luminosity  
with 8-year data takings \cite{Kou:2018nap}. 
An upgrade with five times more luminosity 
is also anticipated with Belle II \cite{Belle-II-upgrade}. 
The BESIII detector is located at the Beijing Electron–Positron Collider II (BEPCII)
with the beam energy ranging from 1.0 GeV to 2.3 GeV 
and luminosity of $10^{33}$ cm$^{-2}$ s$^{-1}$ \cite{Asner:2008nq}. 
STCF is a proposed 
experiment which collides electron with positron at 
the center-of-mass energies in the range 2-7 GeV, 
with the peak luminosity 
${\cal O}(10^{35})$ cm$^{-2}$ s$^{-1}$ 
at 4 GeV \cite{Peng:2019}. 
An integrated luminosity up to 20 ab$^{-1}$ 
is expected to be accumulated with a 10-year STCF runnings, 
assuming 9-month running time each year 
and 90\% data taking efficiency \cite{Peng:2019}. 
The BaBar detector is  
operated at the PEP-II $e^+ e^-$ collider 
from 1999 to 2008 with 
most data collected near  
$\sqrt{s} = 10.58$ GeV 
(the $\Upsilon(4S)$ resonance) 
\cite{Lees:2013rw}.

There are two types of monophoton backgrounds:   
irreducible background and reducible background. 
The irreducible monophoton background is the SM final state containing 
one photon and two neutrinos; 
{one of the irreducible background processes is shown in  
Fig.\ (\ref{F-FeynDia-eeXXA}).} 
The reducible monophoton background arises when 
a photon is produced in the final state together 
with several other visible particles which are 
however not detected due to the limitations of 
the detector acceptance. 
Belle II and BaBar have asymmetric detectors; 
BESIII and STCF have symmetric detectors. 
We discuss the reducible BG in detail later 
for each experiment, since it strongly depends on the 
angular coverage of the detectors.

\section{Belle \uppercase\expandafter{\romannumeral2}}

In Belle II, photons and electrons can be detected in the 
Electromagnetic Calorimeter (ECL), 
which consists of three segments:
forward endcap with $12.4^{\circ} < \theta < 31.4^{\circ}$,
barrel  with $32.2^{\circ} < \theta < 128.7^{\circ}$, 
and backward endcap with $130.7^{\circ} < \theta < 155.1^{\circ}$ 
in the lab frame \cite{Kou:2018nap}.
There are two important monophoton 
reducible backgrounds \cite{Kou:2018nap}: 
one is mainly due to the lack of polar angle  coverage 
of the ECL near the beam directions, which is referred to 
as the ``bBG''; 
the other one is {mainly} due to the gaps between 
the three segments in the ECL detector,  
which is referred to as the ``gBG''.

In the bBG, all the final state particles except the detected monophoton 
are emitted with 
$\theta>155.1^{\circ}$ or $\theta<12.4^{\circ}$ in the lab frame. 
Some major bBG processes include \eesasaa and 
\eeslsla where $\ell = e, \mu, \tau$; 
the final state particles with a slash 
on the name are emitted along the beam directions.

\begin{figure}[htbp]
\begin{center}
\includegraphics[width=0.45\textwidth]{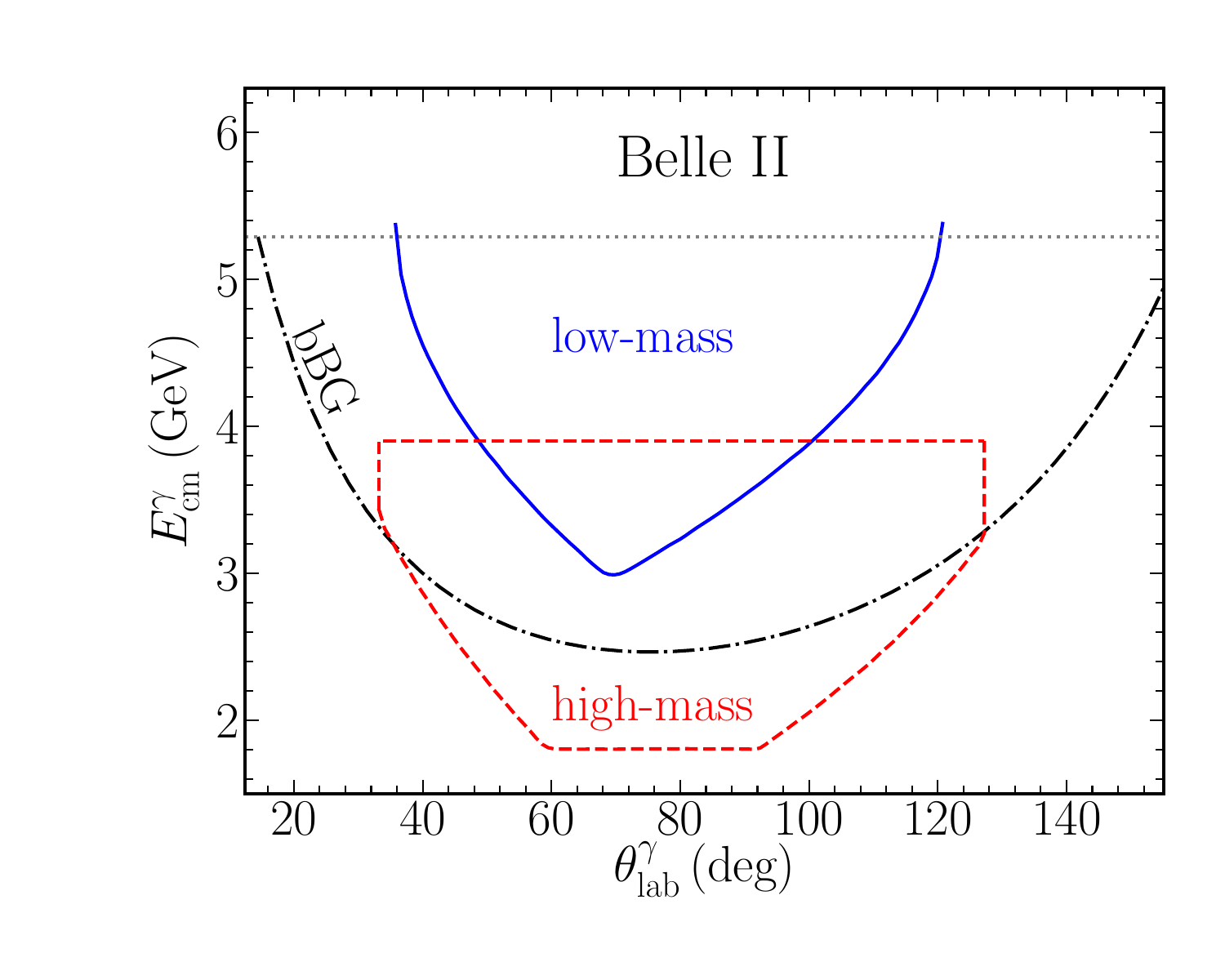}
\caption{Monophoton phase space 
$E_{\rm cm}^\gamma-\theta^\gamma_{\rm lab}$ in Belle-II. 
$E_{\rm cm}^\gamma$ is the photon energy in 
the CM frame; 
$\theta^\gamma_{\rm lab}$ is the photon 
polar angle with respect to  
the initial electron in the lab frame. 
We refer to the region above the blue solid line as  
``low-mass'' region and 
the region enclosed by the red dashed line as 
the ``high-mass'' region. 
Both the ``low-mass''  and ``high-mass''  regions are taken from 
simulations given in Ref.\ \cite{Kou:2018nap}. 
The gray dotted line indicates the $\sqrt{s}/2$ value in Belle-II. 
The black dot-dashed line is the bBG cut. 
}
\label{Fig-belle2cut}
\end{center}
\end{figure}

For symmetric detectors, such as 
BESIII and STCF, 
the maximum energy of the monophoton events 
in the bBG in the CM frame, $E_\gamma^m$, 
is given by  
\be
E_\gamma^m (\theta_\gamma) = 
\sqrt{s} \left(1+{\sin \theta_\gamma \over \sin \theta_{b}}\right)^{-1}, 
\label{Eq-symbBG}
\ee
where $\theta_{b}$ is 
the polar angle 
corresponding to 
the edge of the detector \cite{Liu:2019ogn}. 
For the Belle II detector, which is asymmetric, 
$E_\gamma^m$ in the CM frame is given by 
(if not exceeding $\sqrt{s}/2$) 
\be
E_\gamma^m(\theta_\gamma) = 
\frac{ \sqrt{s}(A\cos\theta_1-\sin\theta_1)}
{A(\cos\theta_1-\cos\theta_\gamma)-(\sin\theta_\gamma+\sin\theta_1)},
\label{eq:bBG}
\ee 
where  
all angles are given in the CM frame, 
and $A=(\sin\theta_1-\sin\theta_2)/(\cos\theta_1-\cos\theta_2)$, 
with $\theta_1$ and $\theta_2$ being 
the polar angles corresponding to 
the edges of the ECL detector.\footnote{The 
polar angle in the CM frame is 
related to that in the lab frame via 
$\tan\theta_{\rm cm} = \sin \theta_{\rm lab}/
(\gamma \cos \theta_{\rm lab}- \gamma \beta )$,
where {$\beta=3/11$} and $\gamma=1/\sqrt{1 - \beta^2}$.}
To remove the above bBG, the detector 
cut $E_\gamma > E_\gamma^m$ is used 
(hereafter the ``bBG'' cut), 
which is shown in Fig.\ (\ref{Fig-belle2cut}).

{Because the ECL gaps are 
significantly away from the beam direction,}
the monophoton energy of the gBG 
can be quite large in the central $\theta_\gamma$ region. 
The gBG simulations have been carried out by 
Ref.\ \cite{Kou:2018nap} in searching for an  
invisibly decaying vector boson. 
The dominated gBG is 
$e^+ e^- \to \gamma \slashed{\gamma} \slashed{\gamma}(\slashed{\gamma})$ with 
at least one final state photon emitting through the gaps \cite{Kou:2018nap}. 
For example, one major background arises when 
one final state photon in the process 
$e^+ e^- \to \gamma \slashed{\gamma} \slashed{\gamma}$ 
escapes via the gap between the ECL barrel and the backward endcap,
and the second photon is emitted along the beam direction \cite{Torben:2019}.
Two different sets of detector cuts are 
designed by Ref.\ \cite{Kou:2018nap} 
to optimize the detection efficiency for different 
masses of the vector boson, 
which are shown in Fig.\ (\ref{Fig-belle2cut}). 
The ``low-mass'' region in the monophoton 
phase space has few gBG events, which 
is applied for the vector boson 
with mass less than $6\ {\rm GeV}$. 
However, if the vector boson mass is in the range $6-8$ GeV, 
only low energy photons can be produced in the new physics processes 
so that the ``high-mass'' cut region is preferred.

\begin{figure}[htbp]
\begin{center}
\includegraphics[width=0.45\textwidth]{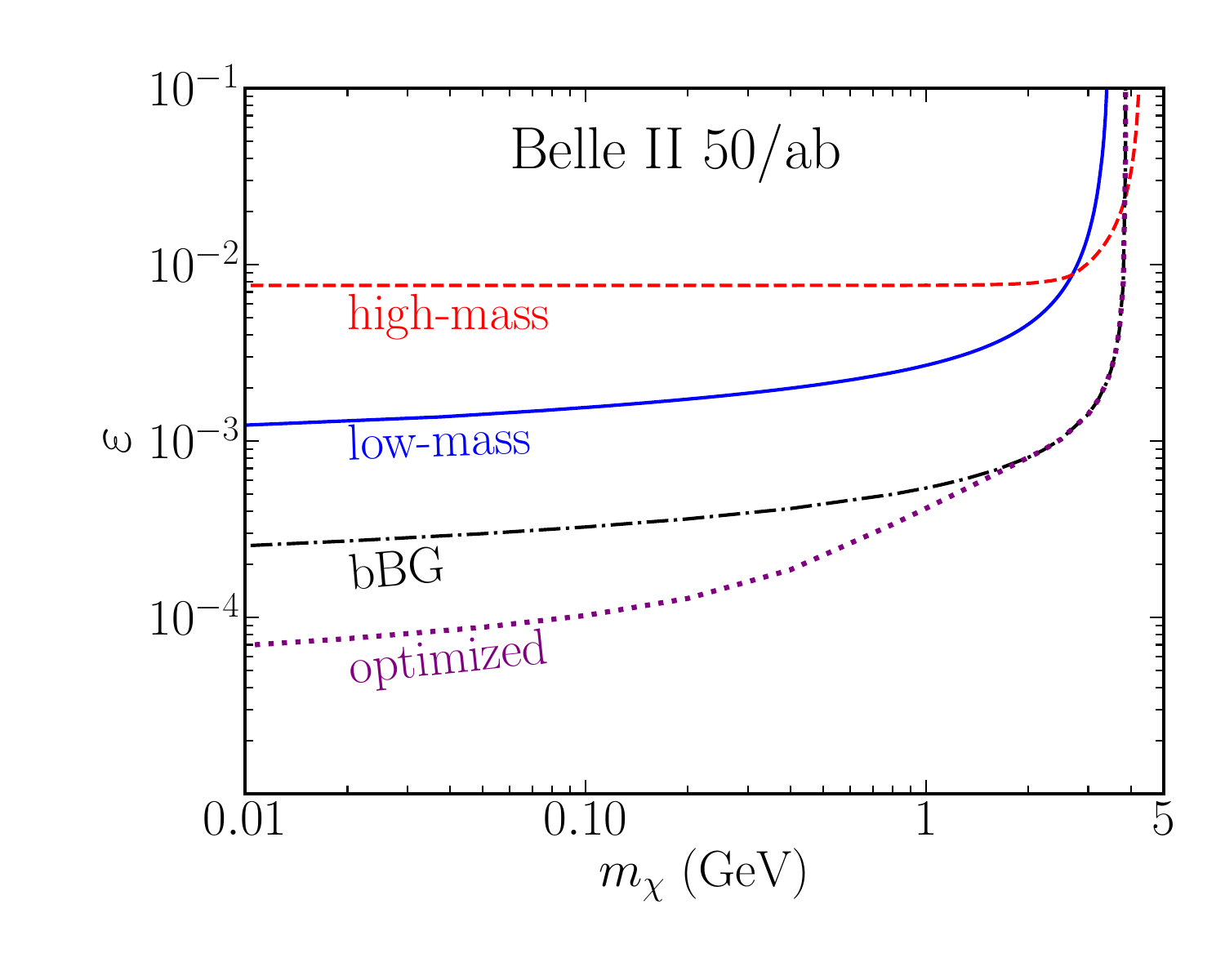}
\caption{The expected 95\% confidence level (C.L.) upper bound 
on millicharge at Belle II under  
the low-mass cut (solid) and 
the high-mass cut (dashed), 
with 50 ab$^{-1}$ integrated luminosity. 
The black (purple) line corresponds to  
the limit using the bBG (optimized) cut. 
}
\label{Fig-Result}
\end{center}
\end{figure}

To probe the millicharge, we define 
$
\chi^{2}(\epsilon) \equiv S^2/(S+B)
$ \cite{Yin:2009mc}, 
where $S$ ($B$) is the number of events in the signal (background) processes. 
The 95\% confidence level (C.L.) upper bound  
on the millicharge, $\epsilon_{95}$, is obtained by solving 
$
\chi^{2} (\epsilon_{95})-\chi^{2} (0)=2.71. 
$
Fig.\ (\ref{Fig-Result}) shows the expected 
95\% C.L.\ upper bound on millicharge using 
the ``low-mass'' and ``high-mass'' cuts 
with 50 $\rm{ab}^{-1}$ data. 
We calculate the signal 
and irreducible background 
events by integrating the differential 
monophoton cross sections 
in different regions of the phase space under different 
detector cuts, 
and assuming photon detection efficiency 
as 95\% \cite{Kou:2018nap}.
Our calculation shows that
there are 
{about 10900 (2280, 15230)} irreducible BG 
events with the bBG (low-mass, high-mass) cut
with 50 ab$^{-1}$ integrated luminosity. 
For the reducible background, 
it is found that about 300 (25000) gBG events 
survived the low-mass (high-mass) cuts with 
20 fb$^{-1}$ integrated luminosity  \cite{Kou:2018nap}, 
which are rescaled according to the luminosity. 
The constraint with the high mass cut becomes better 
than the low-mass cut when the MCP mass 
exceeds $\sim$3 GeV.

We also compute the limits without 
gBG taking into account, in order 
to compare with other experiments 
where detailed simulations with gBG 
are not available. 
We use the bBG cut to remove the reducible 
background events; 
the BG events survived the bBG cut 
are due to irreducible backgrounds, 
if gBG is not considered. 
{We integrated the monophoton differential 
cross section for MCPs \cite{Liu:2018jdi} and 
for SM irreducible BG \cite{Liu:2018jdi}, with the bBG cut to 
obtain the number of events.} 
The 95\% confidence level (C.L.) 
upper bound analyzed with the bBG cut  
is shown in Fig.\ (\ref{Fig-Result}) 
where gBG is not considered; 
the upper bound is about five times stronger than 
the one when gBG is considered under the low-mass cut, 
for mililcharged particles with mass less than 1 GeV.

\section{BESIII and STCF}

It has been recently proposed to search for MCPs  
in BESIII \cite{Liu:2018jdi}. 
Here we update the BESIII sensitivity by taking into account 
the most recent data:  
1.4 (0.13, 0.5) ${\rm fb}^{-1}$ 
at $\sqrt{s} = $ 3.097 (3.554, 3.686) GeV \cite{BESIII:lum}. 
In BESIII, we have $\cos\theta_b=0.95$, 
taking into account the coverage of 
main drift chamber (MDC), electromagnetic calorimeter (EMC), and time-of-flight (TOF)  \cite{Liu:2018jdi}. 
We adopt the detector cuts for photons by 
BESIII Collaboration (hereafter the pre-selection cuts) 
\cite{Ablikim:2017ixv}: 
$E_\gamma>$ 25 MeV with $| \cos \theta |<0.8$ 
or $E_\gamma>$ 50 MeV with $0.86<| \cos \theta |<0.92$.
We further apply the bBG cut to remove the reducible 
background. 
We compute the number of events under 
both pre-selection cuts and the bBG cut
given in Eq.\ (\ref{Eq-symbBG}). 
We define 
$
 \chi^{2}_{\rm tot} (\epsilon)=\sum_{i} \chi_{i}^{2} (\epsilon) , 
$
where 
$
\chi^{2}_i (\epsilon) \equiv S_i^2/(S_i+B_i) 
$
for each BESIII colliding energy. 
The 95\% C.L.\ upper bound on millicharge from BESIII is 
obtained by demanding 
$ \chi^{2}_{\rm tot} (\epsilon_{95}) = \chi^{2} (0) + 2.71$, 
which is shown in Fig.\ (\ref{Fig-Combine-2}). 

\begin{figure}[htbp]
\begin{center}
\includegraphics[width=0.45\textwidth]{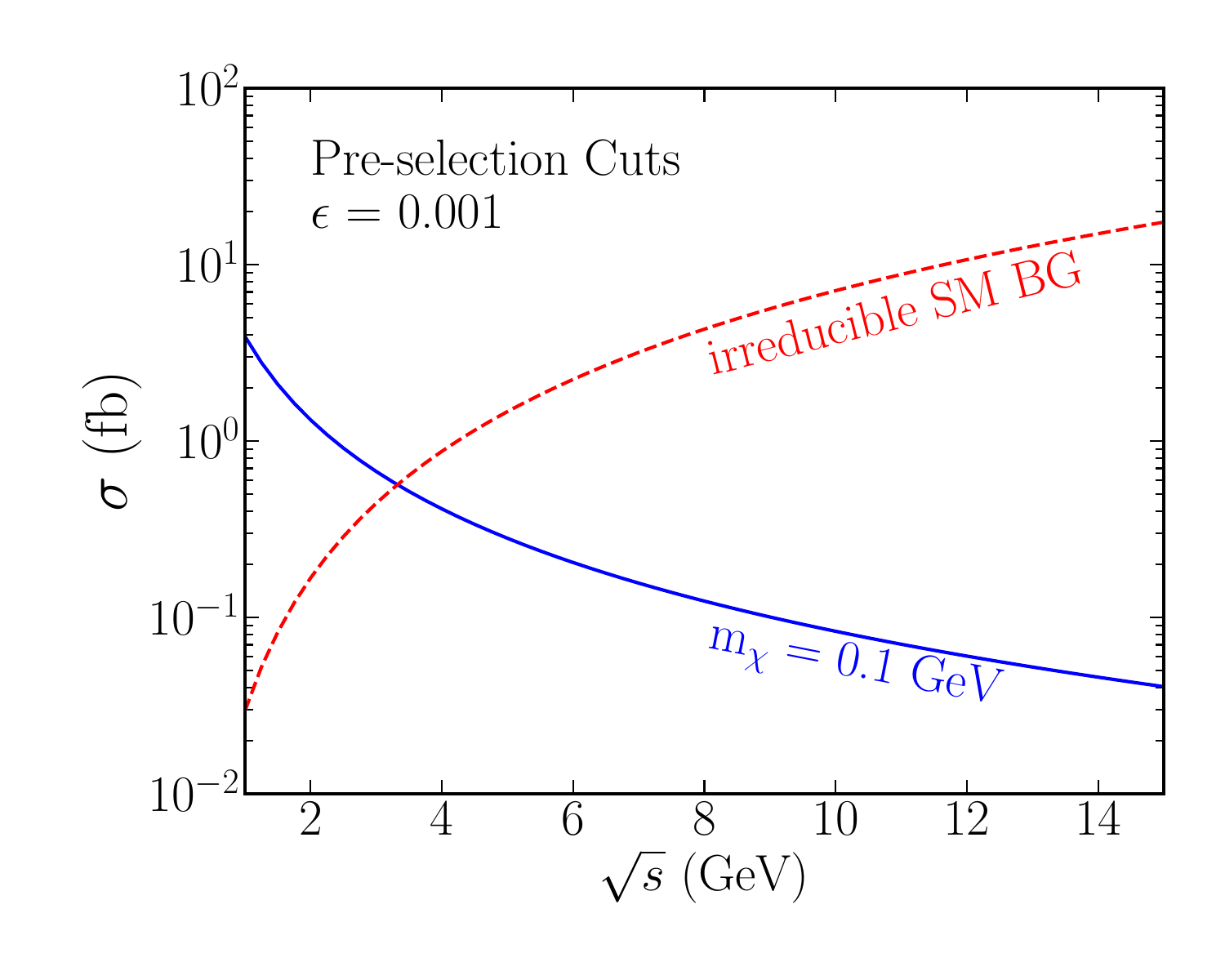}
\caption{Monophoton cross section as a function of 
$\sqrt{s}$ for MCPs (solid), 
and for irreducible BG (dashed). 
Only pre-selection cuts are applied. 
We use $\epsilon=0.001$ and $m_\chi=0.1$ GeV for the MCPs.  
}
\label{Fig-SBCS}
\end{center}
\end{figure}

A total luminosity of 20 ab$^{-1}$ is expected 
at the future STCF experiment operated 
at $\sqrt{s}=2-7$ GeV.
Although the STCF luminosity is a little smaller 
than Belle II, the smaller colliding energy in STCF 
enhances the sensitivity to sub-GeV MCPs.
Fig.\ (\ref{Fig-SBCS}) shows the monophoton cross section 
in the new physics model and in irreducible BG; 
the signal to background ratio increases 
when the colliding energy decreases. 
STCF is thus the ideal experiment to search for light MCPs  
because of both the high integrated luminosity and the relatively 
low colliding energy.

\begin{figure}[htbp]
\begin{center}
\includegraphics[width=0.45\textwidth]{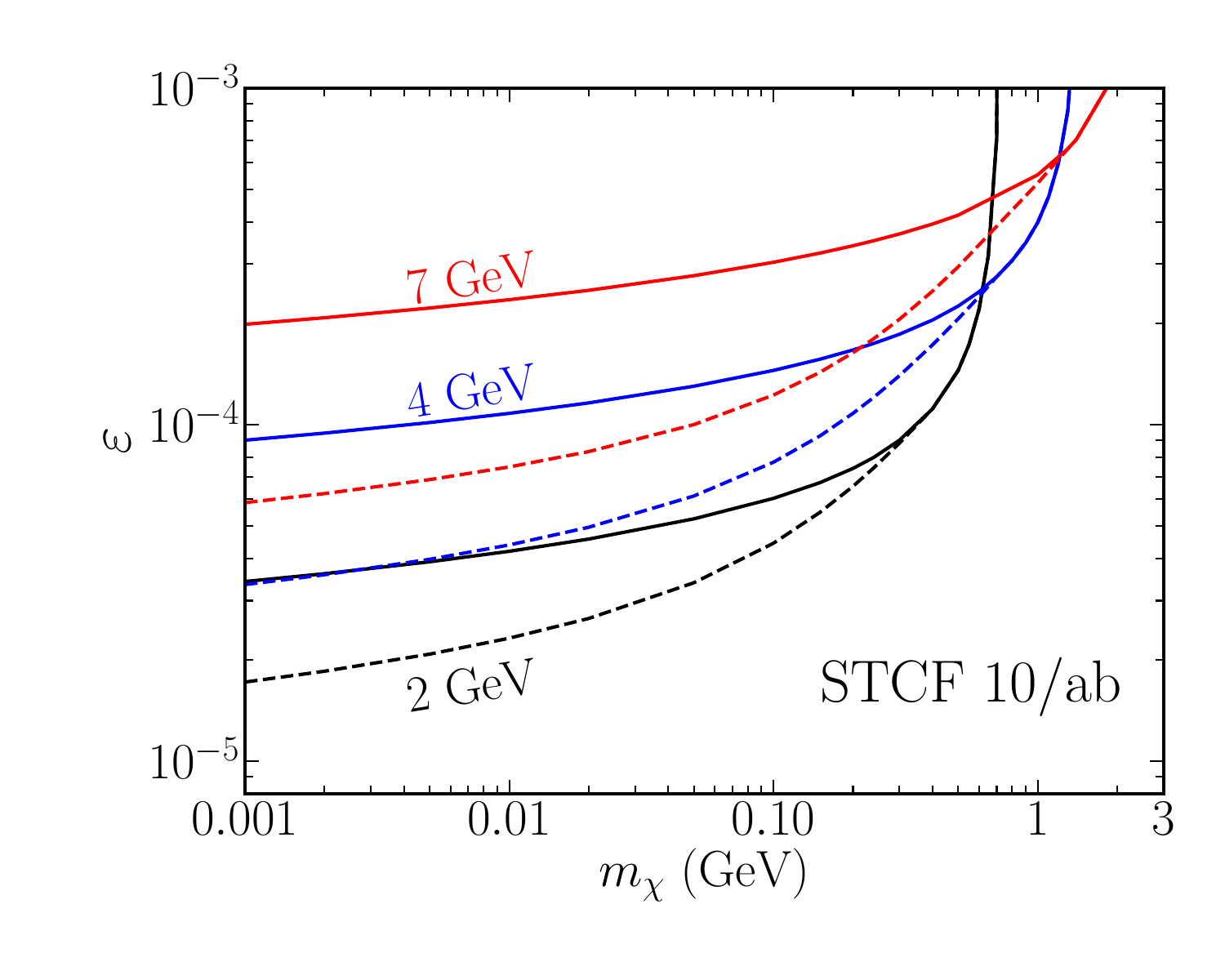}
\caption{The expected 95\% C.L. exclusion limits on millicharge at STCF 
at various colliding energies with 10 ab$^{-1}$ luminosity. 
Solid (dashed) curves indicate the limits under the bBG (optimized) cut 
for $\sqrt{s}=7, 4, 2$ GeV in the descending order. 
}
\label{F-STCF-limit}
\end{center}
\end{figure}

We use the BESIII detector parameters to analyze the 
constraints from STCF, because of the similarity of the 
two experiments. 
Fig.\ (\ref{F-STCF-limit}) shows the expected STCF limits 
on millicharge assuming 10 ab$^{-1}$ luminosity at three 
different colliding energies. 
We compute the signal and irreducible background  
under both the pre-selection cuts and the bBG cut; 
the irreducible BG yields about 27 pb at $\sqrt{s}=4$ GeV 
under these cuts. 
The STCF can probe $\epsilon\simeq 10^{-4}$ 
for mass around 10 MeV, if operated at 
$\sqrt{s}=2$ GeV with 10 ab$^{-1}$ data.

To our knowledge, BESIII has not released 
any analysis on gBG. 
Thus we neglect gBG 
in the BESIII and STCF analyses. 
Improved BESIII and STCF 
limits can be obtained in the 
future when the gBG analysis is available.

\section{BaBar}

To probe MCPs, we use 
the monophoton events collected by 
BaBar collaboration \cite{Aubert:2008as}
which were previously analyzed to search for the light scalar particle $A^{0}$ 
produced via $e^{+} e^{-} \rightarrow \Upsilon(3 S) \rightarrow \gamma A^{0}$; 
two sets of data are analyzed in Ref.\ \cite{Aubert:2008as}: 
the  28 $\rm fb^{-1}$ ``High-E'' photons 
with 3.2 GeV $<E^\gamma_{\rm cm}<5.5$ GeV, 
$-0.31< \cos(\theta^\gamma_{\rm cm}) < 0.6$, 
and $\cos( 6 \phi^\gamma_{\rm cm}) < 0.96$ 
{corresponding to the Instrumented Flux Return (IFR) fiducial};  
the 19 $\rm fb^{-1}$ ``Low-E'' photons 
with 2.2 GeV$<E^\gamma_{\rm cm}<3.7$ GeV 
and $-0.46< \cos (\theta^\gamma_{\rm cm} ) < 0.46$.

The detector cuts used in the BaBar analysis  \cite{Aubert:2008as} 
can be divided into two categories: 
geometric cuts and non-geometric cuts; 
we compute the detector efficiency separately for these 
two cuts, following Ref.\ \cite{Essig:2013vha}.

The detector efficiency for geometric  cuts 
in High-E (Low-E) region is about 34\% (37\%) for the $(1+\cos^2\theta_\gamma)$ 
angular distribution used in Ref.\ \cite{Aubert:2008as}.

Because the total detector efficiency for   
$e^{+} e^{-} \to \Upsilon(3 S) \to \gamma A^{0}$
is (10-11)\% ($20$\%) in the High-E (Low-E) region \cite{Aubert:2008as}, 
the detector efficiency for the non-geometric cuts 
(denoted as ${\rm f}_{\rm NG}$) is about 30\% (54\%) 
in the High-E (Low-E) region. 
The signal events 
under the High-E and Low-E detector cuts is computed via 
\be
 N_{s} = {\cal L}\, f_{\rm NG}  \int d\Omega\, 
d E^d_\gamma\,  dE_\gamma
  f(E^d_\gamma, E_\gamma, \sigma(E_\gamma)) 
\frac{d\sigma}{dE_\gamma dz_\gamma} 
\ee
where 
${d\sigma/(dE_\gamma dz_\gamma)} $ is 
the differential cross section \cite{Liu:2018jdi}, 
$z_\gamma = \cos(\theta_\gamma)$, 
$E^d_\gamma$ is the detected photon energy, 
and $\cal L=$ 28 (19) fb$^{-1}$ for 
High-E (Low-E) data \cite{Aubert:2008as} \cite{Lees:2013rw}. 
Here the photon energy is smeared via the crystal ball function 
$  f(E^d_\gamma, E_\gamma, \sigma(E_\gamma)) $ 
with the energy resolution 
$\sigma(E_{\gamma}) / E_{\gamma} 
=1.5 \% \left(E_{\gamma} / \mathrm{GeV}\right)^{1 / 4} \oplus 1 \%
$ \cite{Essig:2013vha}.

\begin{figure}[htbp]
\vspace{0.2cm}
\centering
\includegraphics[width=0.45\textwidth]{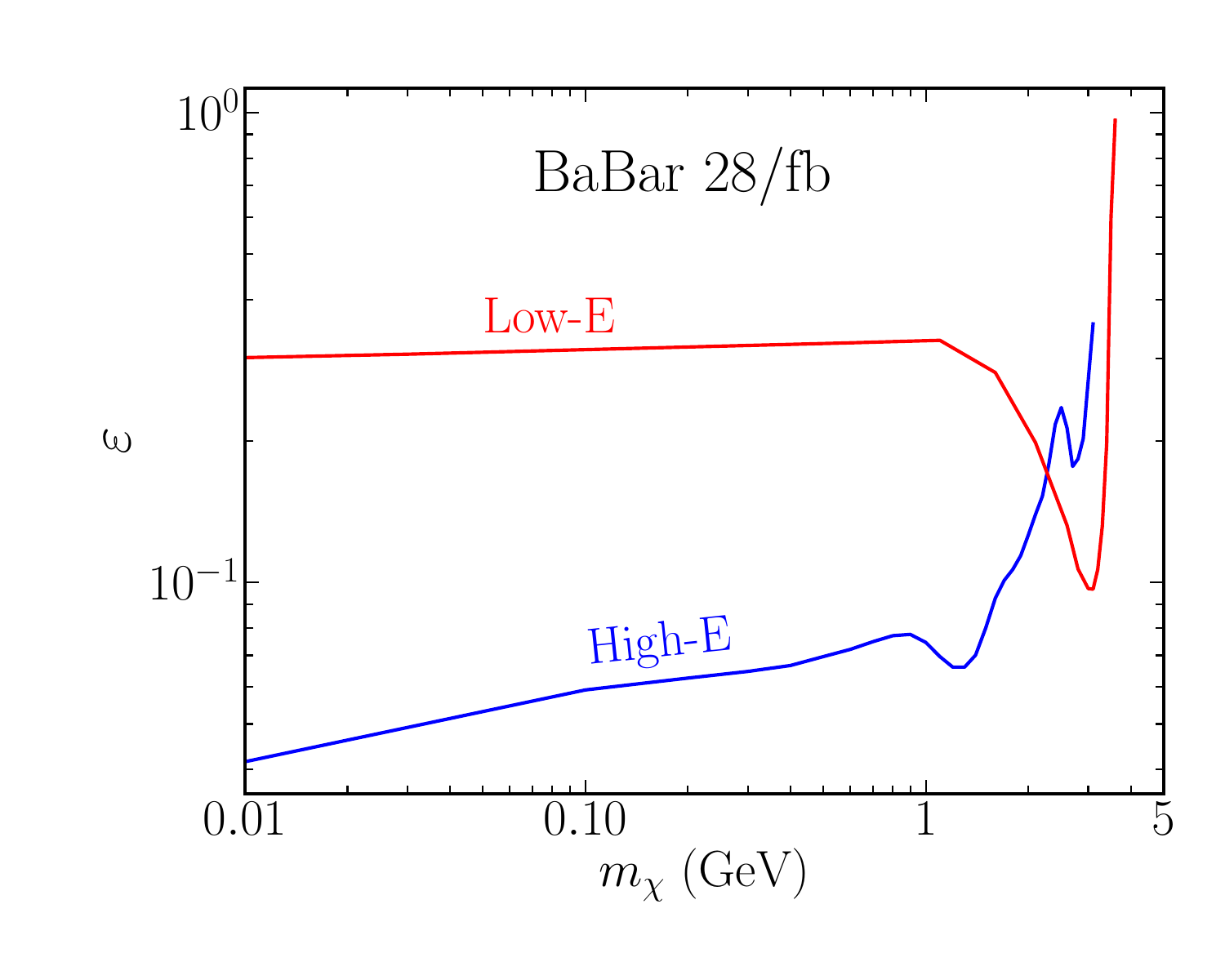}
\caption{The 95\% limits on $\epsilon$ with profile likelihood method at BaBar.
The blue (red) solid line shows the limit with BaBar High-E (Low-E) data.
}
\label{fig-eps95-BABAR}
\end{figure}

Following Refs.\ \cite{Aubert:2008as,Essig:2013vha},
we model the background using fitting functions: 
We use a crystal ball function peaked at $m_{\chi {\chi}} =0$, 
where $m^2_{\chi\chi} = s-2\sqrt{s}E_\gamma$,   
with normalization $N_1$ and $N_{2} \exp(c\,m_{\chi {\chi}}^2)$ 
for the High-E region; 
we use $N_3 \exp(c_1\,m_{\chi {\chi}}^2 + c_2\,m_{\chi {\chi}}^4)$ and  
a constant term $N_4$ 
for the Low-E region. 
The 95\% C.L.\ upper bound on millicharge 
is computed using the profile likelihood method. 
The likelihood function we use is 
\begin{eqnarray}
{\cal L} = {\rm max}\{\prod_{i=1}^{\rm bins}\exp[\frac{(N_s^i+N_b^i-N_o^i)^2}{2 \sigma^2_i}]\},
\end{eqnarray} 
where $N_s^i(N_b^i,N_o^i)$ 
is the number of signal (background, observed) events in bin $i$, 
and {$\sigma_{i}$ is the error bar.} 
We use $N_1, N_2, N_3, c, c_1, c_2$ as nuisance parameters. 
The upper bound on millicharge from the BaBar data 
is shown in Fig.\ (\ref{fig-eps95-BABAR}).

\section{Optimized Cut}

We further carry out a preliminary analysis in which we optimize 
the detector cuts by considering the irreducible background only. 
For the monophoton process at the low energy electron collider, 
 the irreducible background
decreases with photon energy; 
the monophoton cross section 
in the MCP models, however, 
is relatively larger in the 
high energy region than the low energy region.

Thus, selecting photons with relatively high energy 
can enhance discovery sensitivity. 
To find the optimized cut, in addition to the bBG cut, 
we select photons in the range   
$E^{\rm min}_\gamma < E_\gamma < E^{\rm max}_\gamma$,
where $E^{\rm max}_\gamma = (s-4m_\chi^2)/(2\sqrt{s})$ 
and vary $E^{\rm min}_\gamma$ to find the best limit on millicharge. 
Furthermore, the energy difference 
$\Delta E_\gamma \equiv E^{\rm max}_\gamma - E^{\rm min}_\gamma$ 
is required to be larger than the photon energy resolution $\sigma_E$, 
when $E^{\rm max}_\gamma$ is more than 1 $\sigma_E$ above the minimum 
value of the bBG cut curve.

\begin{figure}[htbp]
\begin{center}
\includegraphics[width=0.45\textwidth]{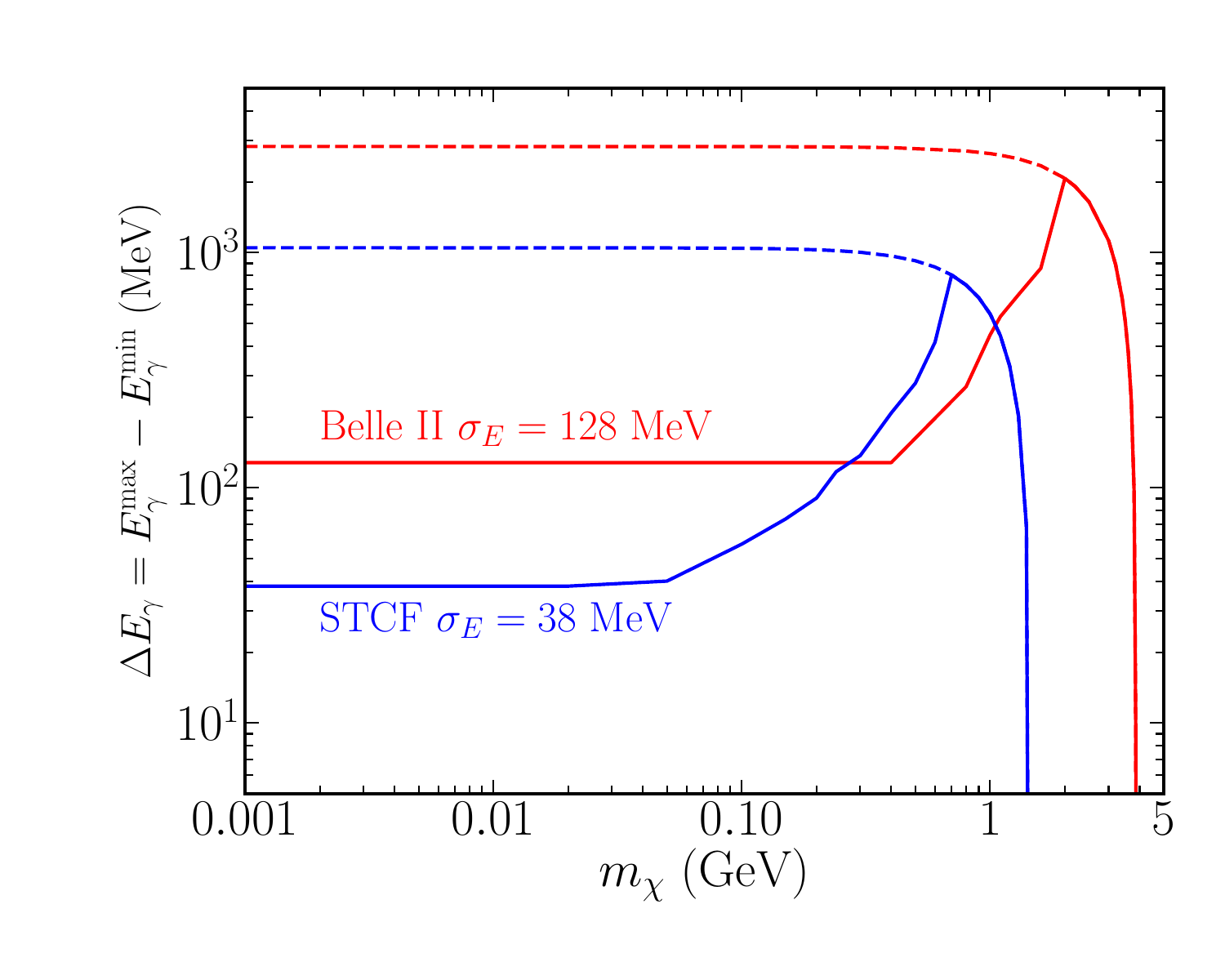}
\caption{The $\Delta E_\gamma$ value that yields 
the best limit on millicharge 
in Belle II and STCF. 
We use 50 (20) $\rm ab^{-1}$ data here 
for Belle II (STCF). 
We consider $\sqrt{s}=4$ GeV for STCF. 
The dashed lines indicate the bBG cuts. 
}
\label{Fig-dE}
\end{center}
\end{figure}

Fig.\ (\ref{Fig-dE}) shows the $\Delta E_\gamma$ that 
gives rise to the best limit on millicharge in Belle II and STCF. 
For STCF, we use the photon resolution of the EMC in BESIII 
$\sigma_E / E=2.3 \% / \sqrt{E / \mathrm{GeV}} \oplus 1 \%$ \cite{Asner:2008nq}, 
and we take $\sigma _{E}=38$ MeV for light mass. 
For Belle II,  $\sigma_E / E = 4\% (1.6\%)$ at 0.1 (8) GeV \cite{Kou:2018nap}
and we take $\sigma _{E}=128$ MeV for light mass.
As shown in Fig.\ (\ref{Fig-dE}), the best $\Delta E_\gamma$ value 
is equal to the photon energy resolution for light mass. 
For high mass, because $E^{\rm max}_\gamma$ starts to approach the 
bBG cut, $\Delta E_\gamma$ can become smaller than 
the photon energy resolution.

\section{Results}

\begin{figure}[htbp]
\begin{center}
\includegraphics[width=0.45\textwidth]{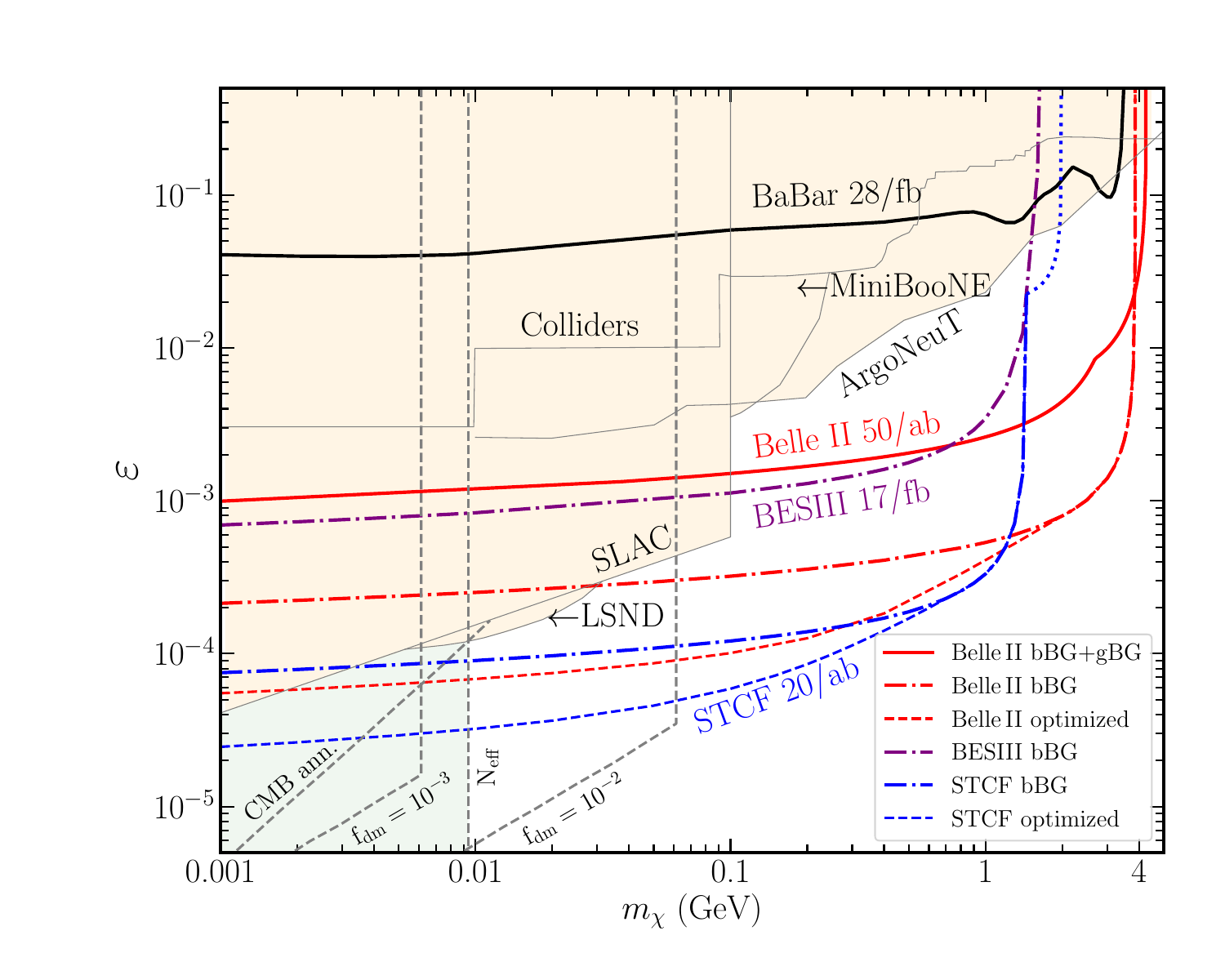}
\caption{The expected 95\% C.L. exclusion limits on MCPs 
at Belle II, BESIII, STCF, and BaBar.
The BaBar limit (black-solid) is obtained by combining the High-E limit and Low-E limit 
in Fig.\ (\ref{fig-eps95-BABAR}). 
The Belle II limit (red-solid) combines the low-mass and high-mass limit in Fig.\ (\ref{Fig-Result}),
where both the bBG and the gBG are considered. 
The other two Belle II limits (red-dot-dashed, red-dashed) 
are obtained with the (bBG, optimized) cuts 
where the gBG is omitted. 
%
The BESIII limit (purple-dot-dashed) is obtained 
with the integrated luminosity during 2011-2018 
where the gBG is omitted.
The STCF limits (blue-dot-dashed, blue-dashed) 
are obtained for $\sqrt{s}=4{\rm\, GeV}$ and $20\ {\rm ab^{-1}}$ 
under the (bBG, optimized) cuts, 
where the gBG is omitted. 
%
Beyond mass $\sim 1.4$ GeV, we analyze monophoton signal  
using the pre-selection cuts to obtain the STCF limit, 
which is shown as the blue-dotted curve. 
Constraints 
from colliders \cite{Davidson:2000hf} \cite{Haas:2014dda}, 
SLAC \cite{Prinz:1998ua}, 
LSND \cite{Magill:2018tbb}, 
MiniBooNE \cite{Magill:2018tbb}, 
and ArgoNeuT \cite{Acciarri:2019jly} 
are also presented.
The parameter space of millicharged DM 
to explain the EDGES 21 cm anomaly \cite{Bowman:2018yin} 
is also shown 
where $f_{\rm dm} = 10^{-3}\ (10^{-2})$ 
is the millicharged DM fraction \cite{Munoz:2018pzp}. 
The excluded regions from cosmology are also shown,
including the bounds from $\rm N_{eff}$  \cite{Berlin:2018sjs, Boehm:2013jpa}
and the dark matter annihilation impact on CMB with  
$f_{\rm dm} = 10^{-2}$  \cite{Slatyer:2015jla, Berlin:2018sjs}.
}
\label{Fig-Combine-2}
\end{center}
\end{figure}

Fig.\ (\ref{Fig-Combine-2}) summarizes the sensitivity on
millicharge $\epsilon$ from the low energy electron 
colliders, including Belle II, STCF, BESIII, and BaBar. 
The BaBar and Belle II limits, 
shown as solid curves on Fig.\ (\ref{Fig-Combine-2}), 
have been analyzed 
taking into account the various SM backgrounds.  
With existing data from BaBar, the previously 
allowed parameter space with millicharge 
$\epsilon \sim 10^{-1}$ and mass $\sim(1-3)$ GeV
can be probed. 
Due to the higher luminosity expected at Belle II, 
a larger parameter space 
that is previously unconstrained by other experiments
is going to be explored by the Belle II;  
with 50 ab$^{-1}$ data, millicharge down to $\sim 10^{-2} - 10^{-3}$ 
for mass $\sim (0.1-4)$ GeV is expected to be probed by Belle II.

The STCF and BESIII limits, 
shown as dot-dashed curves on Fig.\ (\ref{Fig-Combine-2}), 
are obtained when the background due to the gaps in the 
detectors are neglected.
BESIII can probe new parameter space 
for mass $> 100$ MeV, with 17 fb$^{-1}$ data 
collected during 2011-2018. 
The future STCF can probe millicharge parameter space 
below the SLAC experiment \cite{Prinz:1998ua}. 
With 20 ab$^{-1}$ data at $\sqrt{s}=4$ GeV, 
STCF can provide leading constraints on millicharge, 
$\epsilon \lesssim {\cal O}(10^{-4})$ 
for mass from 3 MeV to about 1 GeV.  
The expected limit from STCF also eliminates 
some regions of the 
MCP parameter space where the 21 cm 
anomaly could be explained due to cooling 
from millicharged DM.

In addition to the initial state radiation process in Fig.\ (\ref{F-FeynDia-eeXXA}), 
the $\chi\chi\gamma$ final state can also occur in meson decays 
which can improve the sensitivity for the low mass region. 
However, this is beyond the scope of this work. 
Under the bBG cut, 
STCF loses sensitivity to MCPs when $m_\chi \gtrsim$ 1.5 GeV, 
since  $E^{\rm max}_\gamma = (s-4m_\chi^2)/(2\sqrt{s})$ 
is now lower than the minimum energy of the bBG cut. 
To estimate the STCF sensitivity for $m_\chi \gtrsim$ 1.5 GeV,
we only apply the pre-selection cuts; 
the dominant BG now is due to the \eesesea process. 
The STCF limit in the high mass region 
is shown as the blue dotted curve in Fig.\ (\ref{Fig-Combine-2}).

The omission 
of the gBG in BESIII (17 fb$^{-1}$) 
leads to a stronger limit 
than BelleII (50 ab$^{-1}$) with gBG included 
for $m \lesssim 0.7$ GeV.
To compare the capability of probing the 
parameter space from different experiments, 
we also present a Belle II limit (dot-dashed curve) with gBG omitted.
Although the STCF luminosity is lower than Belle II, 
STCF has better sensitivity in probing the low mass region 
($m\lesssim 1$ GeV) than Belle II.  
This is because STCF is operated at a lower colliding 
energy where the monophoton cross section in 
MCPs (SM) is larger (smaller) than Belle II. 
The one order of magnitude difference in sensitivity 
between the two Belle II limits, 
the solid curve and the dot-dashed curve 
in Fig.\ (\ref{Fig-Combine-2}),  
shows that the control on gGB is very important in 
probing the MCP parameter space. 
Since the dot-dashed curves in Fig.\ (\ref{Fig-Combine-2}) 
are obtained without gBG, the actual limits should be weaker 
when gBG is taken into account. 
However, if the reducible background due to gaps in the detector 
can be significantly suppressed in the future STCF experiment, 
for instance with 
a new sub-detector that can detect the particles emitting 
from the gaps in ECL, 
the one order of magnitude increase in sensitivity from Belle II 
to STCF could be achieved. 
We further computed the 
limits with the optimized detector cuts, 
shown as dashed curves in Fig.\ (\ref{Fig-Combine-2}). 
The optimized detector cuts can further enhance the 
sensitivity of STCF and Belle II in probing the low mass region.

\section{Summary} 

In this paper, we analyzed the sensitivity to 
millicharged particles from four different electron colliders 
operated at the GeV scale: 
BaBar, Belle II, BESIII, and STCF. 
By reanalyzing the 28 fb$^{-1}$ 
monophoton data collected by BaBar, 
one is able to eliminate some 
currently allowed millicharge parameter space 
for $\sim$(0.5-3.5) GeV mass. 
The BESIII experiment can probe 
an even larger region of parameter space 
than BaBar, owing to the lower colliding energy. 
The expected limit on MCPs from BESIII 
is near $\epsilon \sim 10^{-3}$ for 100 MeV mass. 
Projected limits with Belle II and STCF 
experiments are also analyzed. 
It is found that Belle II can probe millicharge 
down to $\epsilon \sim 10^{-3}-10^{-2}$ for 
$0.1$ GeV $\lesssim m \lesssim 4$ GeV. 
The future STCF can further improve the 
sensitivity to low mass MCPs than Belle II 
because it is operated at lower energy. 
Millicharge $\epsilon \lesssim {\cal O}(10^{-4})$ 
for mass from 3 MeV to about 1 GeV 
can be probed by the future 
STCF experiment; 
this excludes 
some of the parameter space for explaining the 21 cm anomaly. 
The sensitivities computed for BESIII and STCF 
are obtained without taking into account the 
gap backgrounds. The more accurate limits 
require full detailed detector simulations, 
which is beyond the scope of this work.

{\it Note added.}---
After the submission of the 
first version of our paper, 
a new limit from ArgoNeuT 
\cite{Acciarri:2019jly} appeared 
which rules out most of the parameter 
space excluded by the BaBar data.

\section{Acknowledgement}

We thank {Shenjian Chen, 
Samuel D. McDermott,
Van Que Tran, 
Jingjing Xu, 
and Lei Zhang} 
for discussions 
and correspondence. 
The work is supported in part  
by the National Natural Science Foundation of China under Grant Nos.\ 
11775109, U1738134, and 11805001.

\appendix

\section{Crystal ball function}

The normalized crystal ball function is given by \cite{book:crystal} 
\begin{align}
\label{eq-crystal}
{f(x , \overline{x}, \sigma)} 
=N  \left\{\begin{array}{ll}{\exp \left(-\frac{(x-\overline{x})^{2}}{2 \sigma^{2}}\right),} & {\text { for } \frac{x-\overline{x}}{\sigma}>-\alpha} \\ {A \left(B-\frac{x-\overline{x}}{\sigma}\right)^{-n},} & {\text { for } \frac{x-\overline{x}}{\sigma} \leqslant-\alpha}\end{array}\right.
\end{align}
where
\begin{align*} 
A &=\left(\frac{n}{|\alpha|}\right)^{n}  \exp \left(-\frac{|\alpha|^{2}}{2}\right), \\ 
B &=\frac{n}{|\alpha|}-|\alpha|, \\ 
N &=\frac{1}{\sigma(C+D)}, \\ 
C &=\frac{n}{|\alpha|}  \frac{1}{n-1}  \exp \left(-\frac{|\alpha|^{2}}{2}\right), \\ 
D &=\sqrt{\frac{\pi}{2}}\left(1+\operatorname{erf}\left(\frac{|\alpha|}{\sqrt{2}}\right)\right). 
\end{align*}
We use $\alpha=\text{0.811}$ and $n=\text{1.79}$ for BaBar  \cite{Essig:2013vha}.

\section{Maximum monophoton energy in reducible BG} 

The maximum energy of the monophoton occurs 
when both final state $e^{\pm}$ are emitted 
at the boundary of ECL 
and are opposite to the photon in the transverse plane. 
Thus the energy-momentum conservation in the CM frame gives rise to 
\bea
&&E_\gamma^{m} \sin\theta_\gamma - E_1 \sin\theta_1 - E_2 \sin\theta_2 = 0\\
&&E_\gamma^{m} \cos\theta_\gamma +E_1 \cos\theta_1+ E_2 \cos\theta_2 = 0 \\
&&E_\gamma^{m} + E_1 + E_2 = \sqrt{s},
\eea
where $E_1$ and $E_2$ are the $e^{\pm}$ energies, 
and $\theta_1$ and $\theta_2$ are the polar angles 
corresponding to the boundary of the ECL. 
The solution for $E_\gamma^{m}$ from the above equations 
yields Eq.\ (\ref{eq:bBG}). 
Note that the monophoton energy cannot exceed $\sqrt{s}/2$.

\end{document}